\documentclass[english,aip,jmp,amssymb,amsmath,reprint]{revtex4-1}
\usepackage[T1]{fontenc}
\usepackage{epstopdf}
\usepackage{amsmath}
\usepackage{amssymb}
\usepackage{mathrsfs}
\usepackage{graphicx}
\usepackage{esint}
\usepackage{babel}
\usepackage{color}
\begin{document}
\title{Thermoelectric-induced unitary Cooper pair splitting efficiency}
\author{Zhan Cao}
\affiliation{Center for Interdisciplinary Studies $\&$ Key Laboratory for Magnetism and Magnetic Materials of the Ministry of Education, Lanzhou University, Lanzhou 730000, China}
\author{Tie-Feng Fang}
\affiliation{Center for Interdisciplinary Studies $\&$ Key Laboratory for Magnetism and Magnetic Materials of the Ministry of Education, Lanzhou University, Lanzhou 730000, China}
\author{Lin Li}
\affiliation{Department of physics, Southern University of science and technology of China, Shenzhen 518005, China}
\author{Hong-Gang Luo}
\affiliation{Center for Interdisciplinary Studies $\&$ Key Laboratory for Magnetism and Magnetic Materials of the Ministry of Education, Lanzhou University, Lanzhou 730000, China}
\affiliation{Beijing Computational Science Research Center, Beijing 100084, China}
\begin{abstract}
Thermoelectric effect is exploited to optimize the Cooper pair splitting efficiency in a Y-shaped junction, which consists of two normal leads coupled to an $s$-wave superconductor via double noninteracting quantum dots. Here, utilizing temperature difference rather than bias voltage between the two normal leads, and tuning the two dot levels such that the transmittance of elastic cotunneling process is particle-hole symmetric, we find currents flowing through the normal leads are totally contributed from the splitting of Cooper pairs emitted from the superconductor. Such a unitary splitting efficiency is significantly better than the efficiencies obtained in experiments so far.
\end{abstract}
\maketitle
Entanglement\cite{{Schrodinger1935},Einstein1935} is a defining feature of quantum mechanical systems.\cite{Menskii2001} The creation of nonlocal pairwise-entangled quantum states, so-called Einstein-Podolsky-Rosen (EPR) pairs, is essential for quantum information processing,\cite{Zeilinger1998} quantum computation,\cite{Steane1998} quantum cryptography,\cite{Ekert1991} and quantum teleportation,\cite{Bennett1993} or more fundamentally, for testing the violation of Bell's inequality. Experiments with entangled photons are well developed and already offer first application.\cite{Ursin2007} However, entanglement between electrons, the fundamental particles of electronics, is difficult to produce in a controlled way. In solid-state systems, the closest electrical analogue to the high energy photons are spin-singlet Cooper pairs in $s$-wave superconductors, which may be adiabatically split into two entangled electrons with opposite spins. The process of converting a Cooper pair into two electrons in different normal metal contacts is called crossed Andreev reflection (CAR) or Cooper pair splitting (CPS) and can lead to positive current cross-correlations.\cite{Beckmann2004,Cadden2009}

\begin{figure}[t]
\centering
\includegraphics[width=\columnwidth]{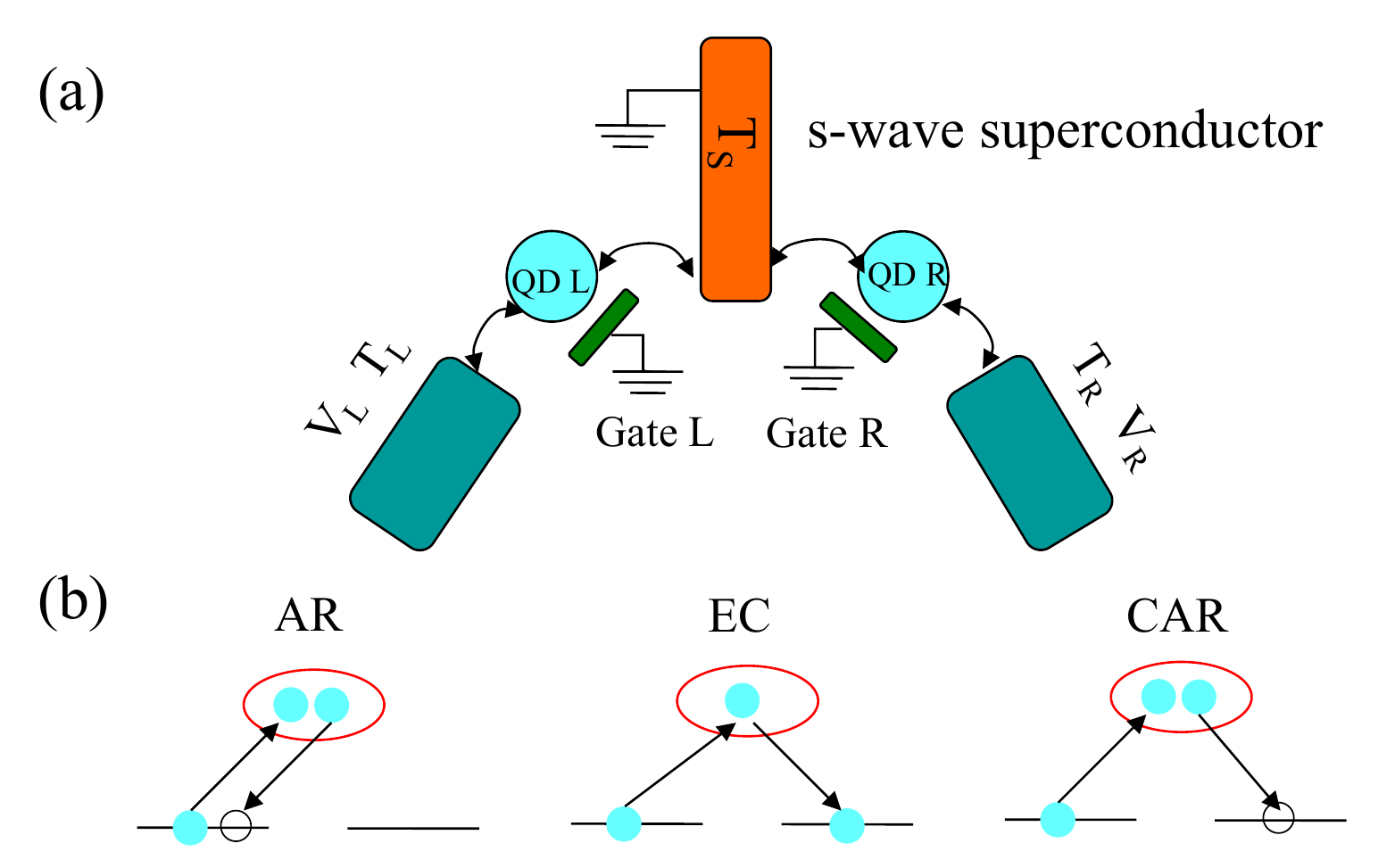}
\caption{(Color online) (a) Schematic view of the Y-shaped three-terminal junction. Two QDs are attached to separate normal leads and coupled to a grounded $s$-wave superconductor. The dot levels can be tuned independently by Gate L and Gate R. Bias and temperature of each lead are indexed. (b) Three elementary transport processes: an electron emitted from one normal lead is reflected back as a hole (AR), transmitted as an electron (EC) or a hole (CAR) into the other lead.}\label{fig1}
\end{figure}

CPS was theoretically proposed to realize in Y-shaped junctions consisting of one superconducting and two normal leads\cite{Torres1999,Recher2001,Chtchelkatchev2002} and has been realized recently in carbon nanotube based setups with\cite{Hofstetter2009,Herrmann2010,Hofstetter2011,Schindele2012,Das2012,Schindele2014} or without\cite{Wei2010} inserting double quantum dots (QDs) between the normal and superconducting leads. In general, the subgap transport occurring in such hybrid structures includes following elementary processes:\cite{Melin2008} an electron emitted from one of the leads is reflected back as a hole, or is transmitted as an electron or a hole into the other lead. The former one is the conventional local Andreev reflection (AR), while the latter two are nonlocal processes which are usually termed as elastic cotunneling (EC) and crossed Andreev reflection (CAR), respectively. Note that the currents flowing through the normal leads contributed from either of the two nonlocal processes can be prohibited, respectively, under simultaneous bias voltage $V_L=V_R$ (no EC contribution) or $V_L=-V_R$ (no CAR contribution).\cite{Floser2013,Cao2015} To promote CPS efficiency, several theoretical proposals have been discussed to suppress the unwanted AR process. For instance, inserting two ferromagnetic metal contacts with antiparallel polarization between the superconductor and respective normal leads blocks the propagation of the opposite spin. Hence the transport of a Cooper pair into a single normal lead is inhibited, while allowing the split pair to pass through the two normal leads.\cite{Lesovik2001,Deutscher2000} However, of course, it requires highly spin-polarized and rotatable ferromagnetic contacts. Recently, splitting efficiency up to $90\%$ has been demonstrated in double QD embedded Y-shaped junctions,\cite{Schindele2012} as sketched in Fig.\,\ref{fig1}(a). Due to the large intra-dot Coulomb interactions, the state with two electrons being on the same QD is strongly suppressed, and thus the split Cooper pair will preferably tunnel into separate dots and subsequently into separate leads.\cite{Schindele2012} Apart from this, it is also proposed to optimize the splitting efficiency by tailoring the time-dependent bias voltages applied to the normal leads.\cite{Pototzky2014} Despite these progresses, for the sake of applications and the explicit demonstration of entanglement, efficiencies close to unity are highly desired.

In this work, we find that in the double QD embedded Y-shaped junctions,\cite{Schindele2012} a unitary CPS efficiency, exceeding the $90\%$ efficiency previously measured, \cite{Schindele2012} can be readily achieved, by exploiting the thermoelectric effect instead of the previous pure electrical scheme, \cite{Schindele2012} even without involving the complicated many-body phenomena in the QDs. The setup we considered [Fig.\,\ref{fig1}(a)] can be described by Hamiltonian with the general form $H=H_{\textrm{leads}}+H_{\textrm{central}}+H_{\textrm{tunnel}}$. The first term describes the left ($L$) and right ($R$) normal leads ($\alpha=L,R$), $H_{\textrm{leads}}=\sum_{k,\alpha,\sigma} \varepsilon_{k}c_{k\alpha\sigma}^\dag c_{k\alpha\sigma}$. The second term describes the central region of the system, $H_{\textrm{central}}=\sum_{\alpha,\sigma}\varepsilon_{\alpha}d_{\alpha\sigma}^{\dag}d_{\alpha\sigma}+\sum_{\alpha}\frac{\Gamma_{\alpha S}}{2}(d_{\alpha\uparrow}d_{\alpha\downarrow}
+\textrm{H.c.})+\frac{\Gamma_{LR}^S}{2}(d_{L\uparrow}d_{R\downarrow}-d_{L\downarrow}d_{R\uparrow}+\textrm{H.c.})$, where $\varepsilon_\alpha$ is the discrete level of the $\alpha$ QD, $\Gamma_{LR}^S=\sqrt{\Gamma_{LS}\Gamma_{RS}}$, and $\Gamma_{\alpha S}=2\pi\rho_s\lambda_\alpha^2$, with $\rho_s$ the normal density of state of the superconductor and $\lambda_\alpha$ the coupling between the $\alpha$ dot and the superconducting lead. The latter two terms of $H_\textrm{central}$, arising due to the large superconducting gap limit $\Delta\rightarrow\infty$, are the superconducting correlations induced in the QDs via their tunnel coupling to the $s$-wave superconductor.\cite{Eldridge2010,Sothmann2014} The tunneling between the QDs and the normal leads is $H_{\textrm{tunnel}}=\sum_{k,\alpha,\sigma}( V_{\alpha}d_{\alpha\sigma}^{\dag}c_{k\alpha\sigma}+\textrm{H.c.})$, with $V_\alpha$ being the tunneling matrix element. An electron and/or hole transferring between the $\alpha$ QD and the $\alpha$ lead is described by an effective tunneling rate $\Gamma_\alpha=2\pi\rho|V_\alpha|^2$, where $\rho$ is the normal-lead density of states. As indicated in Fig.\,\ref{fig1}(a), the bias voltage $V_L$ ($V_R$) is applied to the left (right) lead, while the superconductor is grounded. Since we consider the large gap limit in this work, the temperature ($T_S$) of the superconducting lead, which should be much lower than the superconducting transition temperature, becomes irrelevant. Nevertheless, the temperatures ($T_L$, $T_R$) of the two normal leads play important roles in our theory.

Following the scattering-matrix formalism,\cite{Anantram1996} we obtain the steady current, $I_\alpha=\frac{ie}{\hbar}\langle[N_\alpha,H]\rangle$ with $N_\alpha=\sum_{k,\sigma}c_{k\alpha\sigma}^\dag c_{k\alpha\sigma}$, flowing through the $\alpha$ normal lead including contributions from three elementary transport processes [Fig.\,\ref{fig1}(b)] as
\begin{eqnarray}
&&I_\alpha=\frac{2e}{h}\int d\omega\{\underbrace{\vert s_{\alpha\alpha}^{eh}(\omega)\vert^2[f_{\alpha}^{e}(\omega)-f_{\alpha}^{h}(\omega)]}_{\textrm{AR}}\notag\\
&&+\underbrace{\vert s_{\alpha\bar\alpha}^{ee}(\omega)\vert^2[ f_{\alpha}^{e}(\omega)-f_{\bar\alpha}^{e}(\omega)]}_{\textrm{EC}}+\underbrace{\vert s_{\alpha\bar\alpha}^{eh}(\omega)\vert^2[f_{\alpha}^{e}(\omega)-f_{\bar\alpha}^{h}(\omega)]}_{\textrm{CAR}}\},\notag\\
\label{eq1}
\end{eqnarray}
where $\bar\alpha$ denotes the index opposite to $\alpha$ and $f_{\alpha}^{e,h}(\omega)=[\exp(\frac{\omega\mp eV_{\alpha}}{k_{B}T_\alpha})+1]^{-1}$ are the Fermi distributions of the electron ($e$) and hole ($h$) states in the $\alpha$ lead. The elements of the scattering-matrix are connected to the retarded Green's functions (GFs) via the Fisher-Lee relation \cite{Fisher1981} $s_{\alpha\beta}^{mn}(\omega)=i\delta_{\alpha\beta}\delta_{mn}+\sqrt{\Gamma_\alpha\Gamma_\beta}G^r_{\alpha\beta;mn}(\omega)$, $\{m,n\}= \{e,h\}$. The GF matrix follows from the Dyson equation\cite{Sun1999} $G^r(\omega)=[(g^r(\omega))^{-1}-\Sigma^r(\omega)]^{-1}$, with the bare GF $[g^r(\omega)]^{-1}=\textrm{diag}\{\omega-\varepsilon_L,\omega+\varepsilon_L,\omega-\varepsilon_R,\omega+\varepsilon_R\}$ and
the self energy \begin{eqnarray}
\Sigma^{r}\left(  \omega\right)  =\frac{1}{2}\left(
\begin{array}
[c]{cccc}%
-i\Gamma_{L} & \Gamma_{L}^{S} & 0 &\Gamma_{LR}^S\\
\Gamma_{L}^{S} & -i\Gamma_{L} & \Gamma_{LR}^S & 0\\
0 &\Gamma_{LR}^S & -i\Gamma_{R} & \Gamma_{R}^{S}\\
\Gamma_{LR}^S & 0 & \Gamma_{R}^{S} & -i\Gamma_{R}%
\end{array}
\right),\label{eq2}
\end{eqnarray}
defined in the Nambu basis $\{d_{L\uparrow},d^\dag_{L\downarrow},d_{R\uparrow},d^\dag_{R\downarrow}\}$. Accordingly, the GF matrix elements $G^r_{\alpha\beta;ee}(\omega)$, $G^r_{\alpha\beta;eh}(\omega)$, $G^r_{\alpha\beta;he}(\omega)$, and $G^r_{\alpha\beta;hh}(\omega)$ correspond to $\langle\langle d_\alpha;d_\beta^\dag\rangle\rangle^r_\omega$, $\langle\langle d_\alpha;d_\beta\rangle\rangle^r_\omega$, $\langle\langle d_\alpha^\dag;d_\beta^\dag\rangle\rangle^r_\omega$, and $\langle\langle d_\alpha^\dag;d_\beta\rangle\rangle^r_\omega$, respectively, in the familiar Zubarev notation.\cite{Zubarev1960}

\begin{figure}[t]
\centering
\includegraphics[width=\columnwidth]{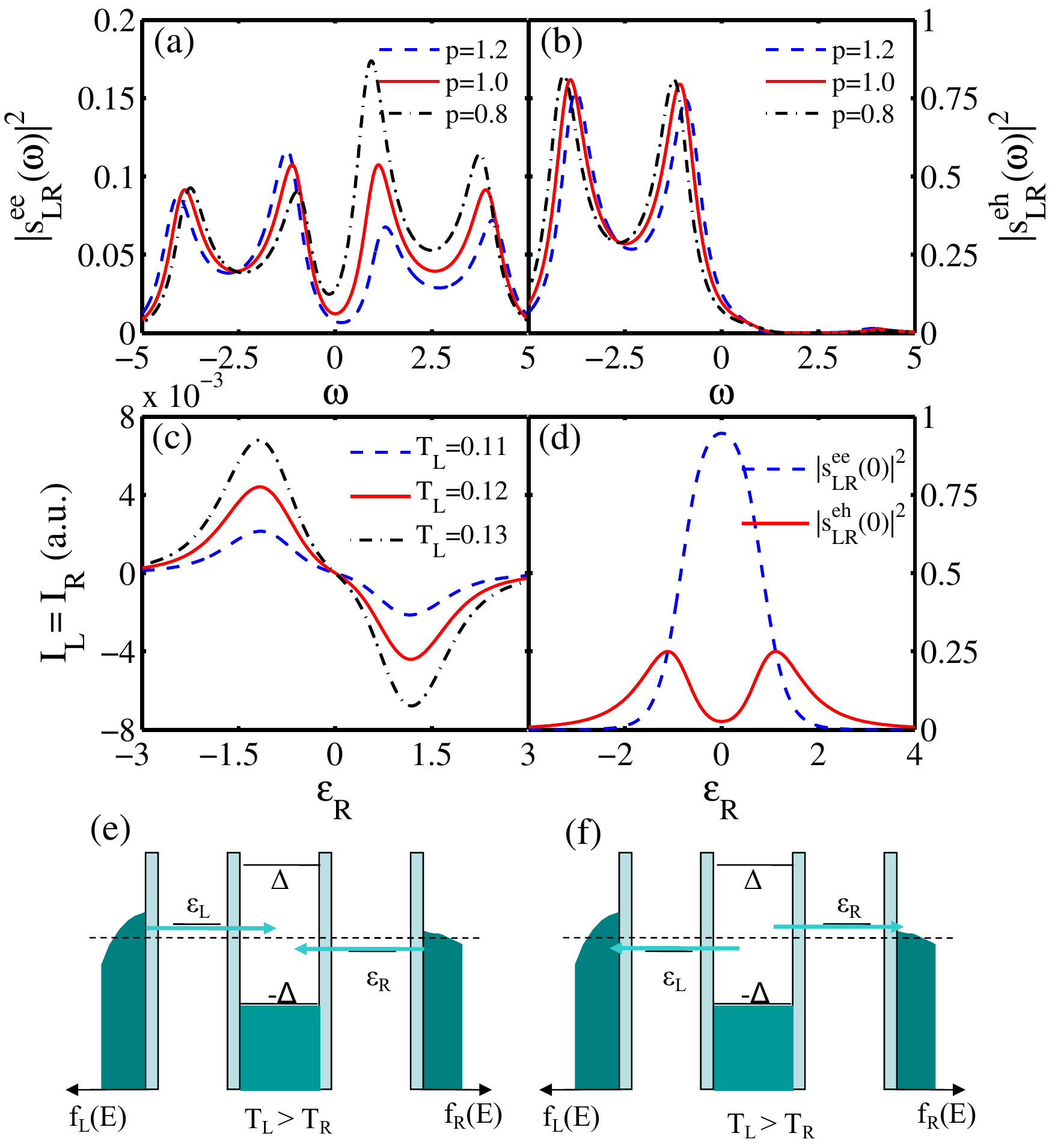}
\caption{(Color online) Energy dependence of (a) EC and (b) CAR transmittance for different $p$ and fixed $\varepsilon_R=2.0$. $p=1$ corresponds to the sweet spot. (c) Total current $I_L=I_R$ contributed only from CAR (i.e., $p=1$) versus $\varepsilon_R$, at different $T_L$ but identical $T_R=0.1$. (d) Zero-energy transmittances of EC and CAR processes versus $\varepsilon_R$ at $p=1.0$. (e) and (f) are representative configurations of dot levels for positive and negative current, respectively. Here $\Gamma_{LS}=\Gamma_{RS}=3$, and $\Gamma_L=\Gamma_R\equiv1$ is used as the unit of energy.}  \label{fig2}
\end{figure}

While Eq.\,(1) describes the general current under arbitrary biases $V_\alpha$ and temperatures $T_\alpha$, we only focus on the thermoelectric effect in this work, i.e., we fix $V_L=V_R=0$ and keep a temperature gradient $T_L\neq T_R$ over the two normal leads. In this case, the electrons and holes have the same distributions in a given lead $f^e_\alpha(\omega)=f^h_\alpha(\omega)$. The AR process is thus projected out of the total currents Eq.\,(1), while the EC and CAR contributions generally survive provided that the temperature $T_L\neq T_R$. To be more precise, the current formula reduces to
\begin{equation}
I_\alpha=\frac{2e}{h}\int d\omega(\vert s_{\alpha\bar\alpha}^{ee}(\omega)\vert^2+\vert s_{\alpha\bar\alpha}^{eh}(\omega)\vert^2)F_{\alpha\bar\alpha}(\omega),\label{eq3}
\end{equation}
with $F_{\alpha\bar\alpha}(\omega)=f_\alpha(\omega)-f_{\bar\alpha}(\omega)$ the difference between the Fermi distributions of the two lead [the index ($e$, $h$) of carrier types is discarded]. Note that $F_{\alpha\bar\alpha}(\omega)$ is an odd-function $F_{\alpha\bar\alpha}(-\omega)=-F_{\alpha\bar\alpha}(\omega)$. It follows that the EC contribution to Eq.\,(\ref{eq3}) would be suppressed exactly if the corresponding transmittance $\vert s_{\alpha\bar\alpha}^{ee}(\omega)\vert^2$ is particle-hole (p-h) symmetric, i.e., an even function of $\omega$. The inserted double QDs provide a controllable way to achieve such p-h symmetry in virtue of the high tunability of the dot levels. Analytically, the p-h symmetric $\vert s^{ee}_{\alpha\bar\alpha}(\omega)\vert^2$ is realized at the \emph{sweet spot} $\Gamma_{LS}/ \Gamma_{RS}=-\varepsilon_L/ \varepsilon_R$, without restriction on the ratio $\Gamma_L/\Gamma_R$. The p-h symmetry of $\vert s^{ee}_{\alpha\bar\alpha}(\omega)\vert^2$ is explicitly illustrated in Fig.\,\ref{fig2}(a), which is broken when $p\equiv -\varepsilon_L/ \varepsilon_R$ slightly deviates from the sweet spot. We note that, for the sake of simplicity, we set $\Gamma_{LS}=\Gamma_{RS}$ for all calculations in this work and hence the sweet spot amounts to $p=1$. Meanwhile, the CAR transmittance is in general asymmetric about the Fermi level [Fig.\ref{fig2}(b)] and thus its contribution to the current never vanishes.

At the sweet spot, the thermoelectric effect results in that a Cooper pair in the superconductor split into two electrons with opposite spins and then tunnel to separate normal leads (negative current), or inversely, two electrons from different normal leads tunnel to the superconductor and then form a Cooper pair (positive current), depending on the dot levels [Fig.\,\ref{fig2}(c)]. These two processes can be explained as follows. Without loss of generality, we assume $T_L>T_R$. There are more (less) electrons being excited above the chemical potential in the left (right) lead, and correspondingly more (less) holes being generated below the chemical potential in the left (right) lead. When $\varepsilon_L>0$ and $\varepsilon_R<0$, an electron with energy $\varepsilon=\varepsilon_L$ in the left lead would like to transfer to the superconductor through dot L and simultaneously a hole is reflected back to the right lead via dot R. This process is equivalent to that two electrons transfer coherently from the two normal leads to the superconductor [Fig.\,\ref{fig2}(e)]. In contrast, the current will flow from the superconductor to the normal leads when $\varepsilon_L<0$ and $\varepsilon_R>0$ [Fig.\,\ref{fig2}(f)]. As the temperature difference between normal leads increases, the amplitude of current is enhanced due to the enhanced unbalance between the left and right Fermi distributions. Moreover, the absolute current changes with the dot level $\varepsilon_R$ non-monotonically, which arrives the peak value near $\varepsilon_R=\pm1$ under the specific parameters we adopted [Fig.\,\ref{fig2}(c)]. In the linear response regime (vanishing temperature difference), the current is approximately determined by the zero-energy CAR transmittance which is maximal at the same dot level position $\varepsilon_R\approx\pm1$ [red solid line in Fig.\,\ref{fig2}(d)].

\begin{figure}[t]
\centering
\includegraphics[width=\columnwidth]{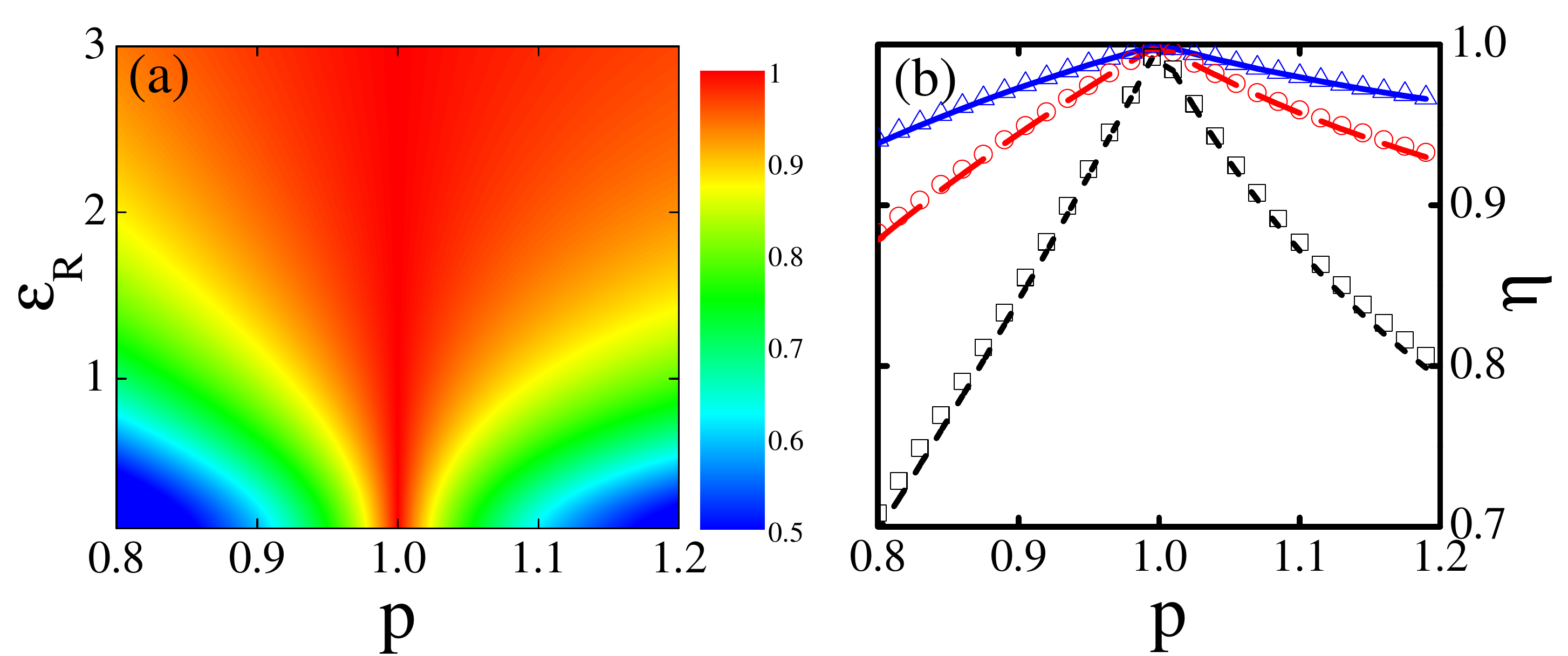}
\caption{(Color online) (a) Density plot of CPS efficiency at $T_L=0.11$. (b) CPS efficiency versus $p$, at $\varepsilon_R=3.0$ (solid line and open triangle), $\varepsilon_R=2.0$ (dashed line and open circle), and $\varepsilon_R=1.0$ (shot-dashed line and open square). Lines and symbols correspond to $T_L=0.11$ and $T_L=0.2$, respectively. Here $T_R=0.1$ and the other parameters are the same as in Fig.\,\ref{fig2}.}  \label{fig3}
\end{figure}

To quantify the dependence of CPS efficiency, \cite{Schindele2012} $\eta\equiv\vert I_\alpha^{CAR}\vert/(\vert I_\alpha^{CAR}\vert+\vert I_\alpha^{EC}\vert)$, on the deviation from the sweet spot, we plot $\eta$ against $p$ for different $\varepsilon_R$ in Figs.\,\ref{fig3}(a) and \ref{fig3}(b). Note that the efficiency defined here is independent of the lead index $\alpha$ since $I_L^{EC}=-I_R^{EC}$ and $I_L^{CAR}=I_R^{CAR}$. It is clear that at the sweet spot ($p=1$) a unitary CPS efficiency $100\%$ is achieved, which prevails over the maximal efficiency of $90\%$ previously measured in experiment.\cite{Schindele2012} Furthermore, as we can see, the CPS efficiency is less sensitive to the deviation from the sweet spot for larger $\varepsilon_R$. This is attributed to the fact that at large $\varepsilon_R$ the EC process is highly blockaded under nearly antisymmetric configuration of dot levels (i.e., $p\sim1$) and thus $\vert I_\alpha^{EC}\vert\ll\vert I_\alpha^{CAR}\vert$. This is consistent with the comparison between the zero-energy transmittance of EC and CAR processes [see Fig.\,\ref{fig2}(d)]. However, we notice that, as $\varepsilon_R$ increases the CAR process will also be suppressed. As a consequence, suitable dot levels should be tuned such that considerable current amplitude as well as the robustness of CPS efficiency against the deviation from the sweet spot are both optimized. Figure \ref{fig3}(b) clearly shows that the CPS efficiency is asymmetric around the sweet spot, which is a direct result of the asymmetric dependence of the transmittances on $p$ [see Figs.\,\ref{fig2}(a) and \ref{fig2}(b)]. Moreover, our CPS efficiency is also found insensitive to the temperature difference between the normal leads [Fig.\,\ref{fig3}(b)].

Although its net contribution to the total currents vanishes at the sweet spot, the EC process does take place in the transport. There is the probability that in the two-particle transport the electron participating the CAR process exchanges role with the one in EC process and vice versa, which would destroy the entanglement between the split electrons entering the separate normal leads. One of the reliable ways to evaluate the entanglement preservation is to study the current cross-correlation between the left and right leads, \cite{Wei2010,Belzig2014} defined as $S_{LR}(t)=\langle \{\delta \hat{I}_{L}(t),\delta \hat{I}_{R}(0)\}\rangle$. At lowest order in the tunneling amplitudes between the normal leads and the superconductor, the two nonlocal processes are decoupled, leading to positive (negative) current cross-correlation for the CAR (EC) process.\cite{Bignon004} A simple interpretation of this fact is that CAR implies instantaneous currents of the same sign in both leads, while EC implies instantaneous currents of opposite signs. Following Ref.\,\onlinecite{Anantram1996}, and taking $V_L=V_R$ and $T_L\neq T_R$, the zero-frequency power spectrum $S_{LR}\equiv\int dtS_{LR}(t)$ of our hybrid structure can be readily obtained $S_{LR}=S_{LR}^{AA}+S_{LR}^{AB}$ with
\begin{eqnarray}
S_{LR}^{AA}=&&-\frac{2e^{2}}{h}\sum_{m}\int d\omega\{  \vert s_{LR}^{mm}(\omega)  \vert ^{2}f_{R}(\omega)  [  1-f_{R}(  \omega)  ]\notag\\
&&+\vert s_{RL}^{mm}(  \omega)  \vert ^{2}f_{L}(\omega) [  1-f_{L}(\omega)] \notag\\
&&+\vert\sum_{\alpha, n}s_{R\alpha}^{mn}(  \omega)  s_{L\alpha}^{mn\dag}(  \omega)  f_{\alpha}(  \omega) \vert^{2}\}, \label{eq4}
\end{eqnarray}
\begin{eqnarray}
S_{LR}^{AB}=&&\frac{2e^{2}}{h}\sum_{m}\int d\omega\{ \vert s_{LR}^{m\bar{m}}(  \omega) \vert ^{2}f_{R}(\omega) [  1-f_{R}(  \omega) ]\notag\\
&&+\vert s_{RL}^{\bar{m}m}(  \omega) \vert ^{2}f_{L}(  \omega) [  1-f_{L}(  \omega)  ]\notag\\
&&+\vert \sum_{\alpha, n}s_{R\alpha}^{\bar{m}n}(  \omega)s_{L\alpha}^{mn\dag}(  \omega)  f_{\alpha}(\omega) \vert ^{2}\}. \label{eq5}
\end{eqnarray}
Here $S_{LR}^{AA}$ accounts for the current cross-correlation between same-type carriers, while $S_{LR}^{AB}$ for that between different-type carriers.

\begin{figure}[t]
\centering
\includegraphics[width=0.9\columnwidth]{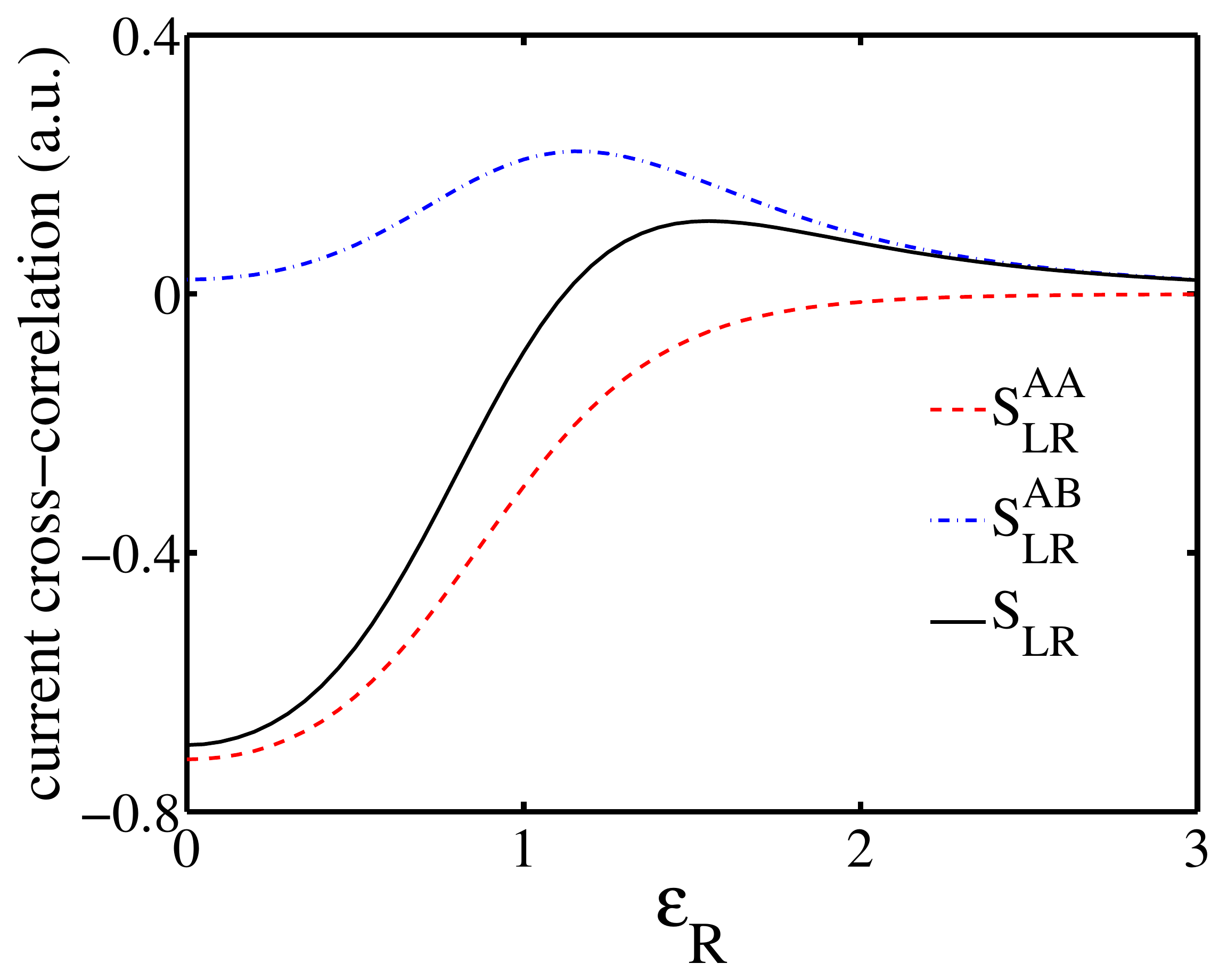}
\caption{(Color online) Dependence of current cross-correlation between same-type carriers ($S_{LR}^{AA}$), different-type carriers ($S_{LR}^{AB}$), and the total cross-correlation ($S_{LR}$) on the dot level $\varepsilon_R$, at the sweet spot. Here $T_L=0.11$ and $T_R=0.1$, the other parameters are the same as in Fig.\,\ref{fig2}.}  \label{fig4}
\end{figure}

In Fig.\,\ref{fig4}, we display the dependence of $S_{LR}^{AA}$, $S_{LR}^{AB}$, and $S_{LR}$ on the dot level $\varepsilon_R$ at the sweet spot. Here $\varepsilon_R$ is restricted to above the Fermi level such that Cooper pairs split into the normal leads. It is shown that $S_{LR}^{AA}$ and $S_{LR}^{AB}$ are always negative and positive, respectively. This property is evident from the structures of Eqs.\,(\ref{eq4}) and (\ref{eq5}), and thus the total cross-correlation $S_{LR}$ can be either positive or negative depending on the relative strengths of the two components. As $\varepsilon_R$ goes away from the Fermi level, $S_{LR}^{AA}$ increases monotonically, while $S_{LR}^{AB}$ evolves firstly to a maximum and then turns to decrease. Moreover, $S_{LR}^{AB}$ approaches to zero more slowly than $S_{LR}^{AA}$, since that $S_{LR}^{AB}$ is almost identical to $S_{LR}$ when $\varepsilon_R$ becomes large enough. This is roughly explained by the dependence of EC and CAR transmittance on the dot level $\varepsilon_R$. The former is intensely suppressed in comparison with the latter when $\varepsilon_R$ exceeds the point at which $\vert s_{\alpha\bar\alpha}^{ee}(\omega)\vert=\vert s_{\alpha\bar\alpha}^{eh}(\omega)\vert$ [see Fig.\,\ref{fig2}(d)]. According to these analysis on the currents and current cross-correlations, in order to obtain high-quality entangled electron pairs in separate normal leads, the dot levels have to be tuned to the sweet spot, and further more to make a compromise between i) considerable current amplitudes, ii) robustness of CPS efficiency against the deviation from the sweet spot, and iii) low enough EC transmittance so that to diminish the $S_{LR}^{AA}$.

In conclusion, we propose utilizing the thermoelectric effect to realize the CPS in a double QD embedded Y-shaped junction, on the contrary to the previous electrical schemes. \cite{Schindele2012} Unitary splitting efficiency, exceeding the previous efficiency \cite{Schindele2012} of $90\%$, is predicted by tuning the dot levels such that the EC transmittance is p-h symmetric. In addition, in order to obtain high-quality electron pairs which preserve the entanglement the dot levels are suggested to be tuned to the situation where the CAR transmittance dominates over the EC transmittance.

This work is supported by NSFC (Grants Nos. 11174115, 11325417), PCSIRT (Grant No. IRT1251).

\end{document}